\documentclass[sigconf]{acmart}

\usepackage{soul}
\usepackage{xcolor}
\definecolor{highlightblue}{HTML}{16B1FF}
\definecolor{highlightgreen}{HTML}{56CA00}
\definecolor{highlightyellow}{HTML}{FFB400}

\AtBeginDocument{%
  }

\setcopyright{acmlicensed}
\copyrightyear{2018}
\acmYear{2018}
\acmDOI{XXXXXXX.XXXXXXX}
\acmConference[Conference acronym 'XX]{Make sure to enter the correct
  conference title from your rights confirmation emai}{June 03--05,
  2018}{Woodstock, NY}

\begin{document}

\title{Augmented Library: Toward Enriching Physical Library Experience Using HMD-Based Augmented Reality}

\author{Qianjie Wei}
\affiliation{%
  \institution{The Hong Kong University of Science and Technology (Guangzhou)}
  \city{Guangzhou}
  \country{China}}
\email{qwei883@connect.hkust-gz.edu.cn}
\orcid{0009-0001-4429-4499}

\author{Jingling Zhang}
\affiliation{%
  \institution{The Hong Kong University of Science and Technology (Guangzhou)}
  \city{Guangzhou}
  \country{China}}
\email{jzhang898@connect.hkust-gz.edu.cn}
\orcid{0009-0006-6555-5201}

\author{Pengqi Wang}
\affiliation{%
  \institution{The Hong Kong University of Science and Technology (Guangzhou)}
  \city{Guangzhou}
  \country{China}}
\email{pwang294@connect.hkust-gz.edu.cn}
\orcid{0000-0002-7221-5348}

\author{Xiaofu Jin}
\affiliation{%
  \institution{The Hong Kong University of Science and Technology}
  \city{Hong Kong SAR}
  \country{China}}
\email{xjinao@connect.ust.hk}
\orcid{0000-0002-7239-3769}

\author{Mingming Fan}
\authornote{Corresponding author}
\affiliation{%
  \institution{The Hong Kong University of Science and Technology (Guangzhou)}
  \city{Guangzhou}
  \country{China}}
\affiliation{%
  \institution{The Hong Kong University of Science and Technology}
  \city{Hong Kong SAR}
  \country{China}}
\email{mingmingfan@ust.hk}
\orcid{0000-0002-0356-4712}

\renewcommand{\shortauthors}{Wei, et al.}

\begin{abstract}
Despite the rise of digital libraries and online reading platforms, physical libraries still offer unique benefits for education and community engagement. However, due to the convenience of digital resources, physical library visits, especially by college students, have declined. This underscores the need to better engage these users. Augmented Reality (AR) could potentially bridge the gap between the physical and digital worlds. In this paper, we present \textit{Augmented Library}, an HMD-based AR system designed to revitalize the physical library experience. By creating interactive features that enhance book discovery, encourage community engagement, and cater to diverse user needs, \textit{Augmented Library} combines digital convenience with physical libraries' rich experiences. This paper discusses the development of the system and preliminary user feedback on its impact on student engagement in physical libraries.
\end{abstract}




\maketitle

\section{Introduction}

Physical libraries play a vital role in education and community engagement by offering distraction-free environments that foster focused reading, accidental discoveries, personal connections with knowledge, and face-to-face social interactions \cite{kim_user_2017, shoham_academic_2019, li_impacts_2018a, bilandzic2014learning, mcdonald2006ten}. However, the convenience of online platforms has led many to prefer digital libraries and reading apps \cite{tyler_factors_2011, silver_are_2005}, raising concerns about the future of physical libraries \cite{cocozza2017ebooks} and highlighting the need for them to incorporate digital technologies to stay relevant.

Prior work has sought to enhance navigation and social interactions in libraries\cite{10.1145/1185448.1185531, bilandzic2014learning}, but often lacks focus on the exploration of physical books, which is crucial to the library experience and its cultural appeal\cite{barclay2017space}. Therefore, this paper aims to design a system centered on physical books to encourage library exploration.

Augmented Reality (AR) offers promising ways to enhance the library experience by overlaying digital content onto the physical world\cite{li2020exploring}. AR can enrich interactions around physical books, encouraging exploration and reinforcing the library's role as a hub for social and intellectual exchange\cite{hirskyj2020social}. In this work, we explored two research questions (RQs): \textbf{RQ1:} What are design considerations for augmenting physical library experience via AR? \textbf{RQ2:} How might an AR-augment library enhance users' experiences?

To answer RQ1, we designed a low-fidelity AR prototype and conducted semi-structured interviews with five librarians. We identified three main challenges. Based on the challenges, we developed \textit{Augmented Library}. Key AR features include color-coded user groups and virtual tags to identify diverse stakeholders, data-driven augmented bookshelves to enhance book discovery, and interactive book reviews to foster community engagement. 

To answer RQ2, we conducted user evaluation with university students (N=10). The results showed that \textit{Augmented Library} enhanced book discovery, with positive feedback on visualization features and recommendations. Participants found it easy to locate books recommended by peers and instructors, though concerns about privacy and the comment feature's usability highlighted areas for improvement. There was also a desire for more interactive elements, such as gamified features and a clearer tagging system.

\section{Related Work}

\textbf{Offline Public Libraries.} Public libraries are crucial for personal development and social interaction \cite{buschman2003dismantling, mccabe2001civic}. They offer extensive print collections, with many studies highlighting the benefits of print reading \cite{larson2010digital, 10.1145/258549.258787}, such as faster reading and better error detection compared to screens \cite{10.1145/2838739.2838745}. 
Modern library designs often feature open spaces or transparent glass partitions \cite{bilandzic2014learning, mcdonald2006ten}, encouraging spontaneous discoveries among people working nearby.
However, with the rise of digital technology, many people, particularly the young, have grown up with easy online access to books \cite{loh2019d}. This led to predictions that traditional books might become obsolete due to online reading and digital tools \cite{cocozza2017ebooks}. Although these predictions have not materialized \cite{cocozza2017ebooks}, libraries still need to embrace and integrate digital technologies into their services to remain relevant and continue to attract and retain users.

\textbf{Enhancing Library Visit Experience.} Technologies have been applied to improve how people \textbf{navigate and explore} libraries. For instance, the "BookMark" study used barcodes on books to create a cost-effective indoor navigation system within libraries \cite{PEARSON201722}. Brian et al. utilized handheld devices to help users find services and resources within the library \cite{10.1145/1185448.1185531}. 
To enhance people's \textbf{social experiences} in libraries, a study used digital wallpapers and public displays to present visitors' background information, thereby fostering communal interaction \cite{bilandzic2014learning}. However, such an approach may divert attention from books, which have been the central focus of library activities for decades \cite{barclay2017space}, potentially diminishing the unique appeal of libraries.

\textbf{AR Promotes Socialization.} Over recent years, significant advancements have been made in the capabilities of AR. Breakthroughs in sensor-based tracking \cite{newcombe2011kinectfusion} and vision algorithms \cite{azinovic2022neural} enable AR devices to seamlessly integrate accurate real-world information with virtual content. Research indicated that social AR can enhance the creation and sharing of information and ideas by overlaying digital content onto the physical world, thereby influencing face-to-face interactions \cite{hirskyj2020social}. For instance, an AR social learning game helped elementary students practice mathematics collaboratively \cite{li2020exploring}, and a mobile AR application was developed for multi-user interactions around art and cultural heritage \cite{10.1145/3582266}. Inspired by prior work, we first designed a low-fidelity prototype and then conducted a formative study to explore how AR could enrich social interactions around books in libraries, encouraging broader exploration and discovery.

\section{Formative Study}

\subsection{Low-fidelity Prototype}

\begin{figure}[htbp]
    \centering
    \includegraphics[width=\columnwidth]{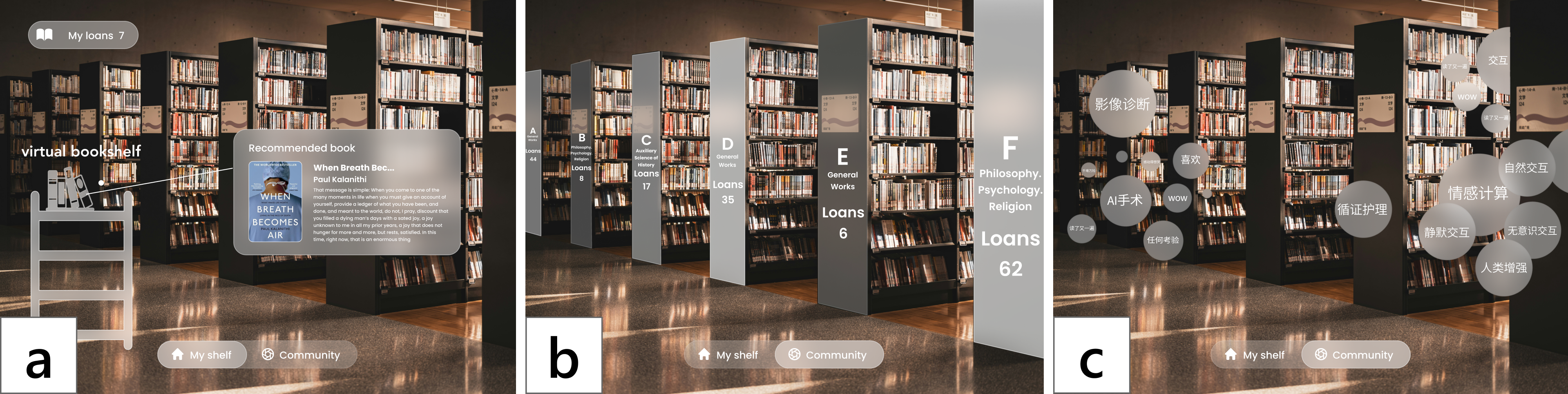}
    \caption{Low-fidelity prototype. (a) AR personal bookshelf. (b) Augmented bookshelves. (c) Comments bubble.}
    \label{fig:Low-fidelityPrototype1}
\end{figure}

Using Figma, we designed a low-fidelity prototype to enhance exploration and engagement in physical libraries. Key AR components included (\autoref{fig:Low-fidelityPrototype1}): 
(1) \textbf{AR Personal Bookshelf.} Integrates online personal data to make library exploration more engaging by presenting customized book recommendations through an AR personal bookshelf.
(2) \textbf{Augmented Bookshelves.} Overlays digital information on bookshelves, such as each category's book loan volumes.
(3) \textbf{Comments Bubble.} Visualizes virtual comments from other readers as AR bubbles near relevant books, fostering a sense of community within the library.

\subsection{Interviewing Librarians}

We invited 5 librarians from a local university library for interviews. Participants ranged in age from 26 to 52 years old (M=35.4), with 2 females and 3 males. Only one of them had AR HMDs experience. We began with a background interview and an AR introduction. Then we conducted a structured interview to explore the unique attributes, advantages, and issues of physical libraries. Finally, we showed a video of our low-fidelity prototype and solicited feedback and suggestions for improvements and additional functionalities.

\subsection{Findings}

Librarians (N=5) appreciated the potential to display book loan volumes on augmented bookshelves. Furthermore, three main challenges were identified:

\textbf{C1: Importance of Enhancing Book Discovery.} This challenge aims to boost the visibility and accessibility of library resources. L2 noted, "\textit{The visual display of book loan volumes will spark curiosity among students.}" L4 mentioned, "\textit{By seeing this AR data, students can clearly know trending books in the library.}" Therefore, the system is suggested to use AR displays to create data visualizations, which would help users discover more resources.

\textbf{C2: Needs for Fostering Community Engagement.} L4 noted, "\textit{Unlike online platforms, library spaces cannot allow people to exchange ideas without time constraints.}" L5 also mentioned, "\textit{Seeing comments pop up next to books can start some meaningful conversations among students.}" This derived that we could stimulate community dialogue and engagement by displaying virtual comments near relevant books.

\textbf{C3: Needs for Identifying Diverse Stakeholders.} Participants pointed out three primary stakeholders of university libraries. Each group has distinct roles and needs. (L1: "\textit{Students are the primary users, utilizing resources for learning and research. Faculty guide students in their academic pursuits. While librarians manage library resources.}") (L2: "\textit{Meeting the specific needs of various library stakeholders requires tailored solutions.}") Both L1 and L2 suggested using different colors to differentiate user groups. - For C3, the AR system should consider the distinct roles and expectations of students, teachers, and librarians.

\textbf{
\begin{figure*}[htbp]
    \centering
    \includegraphics[width=\textwidth]{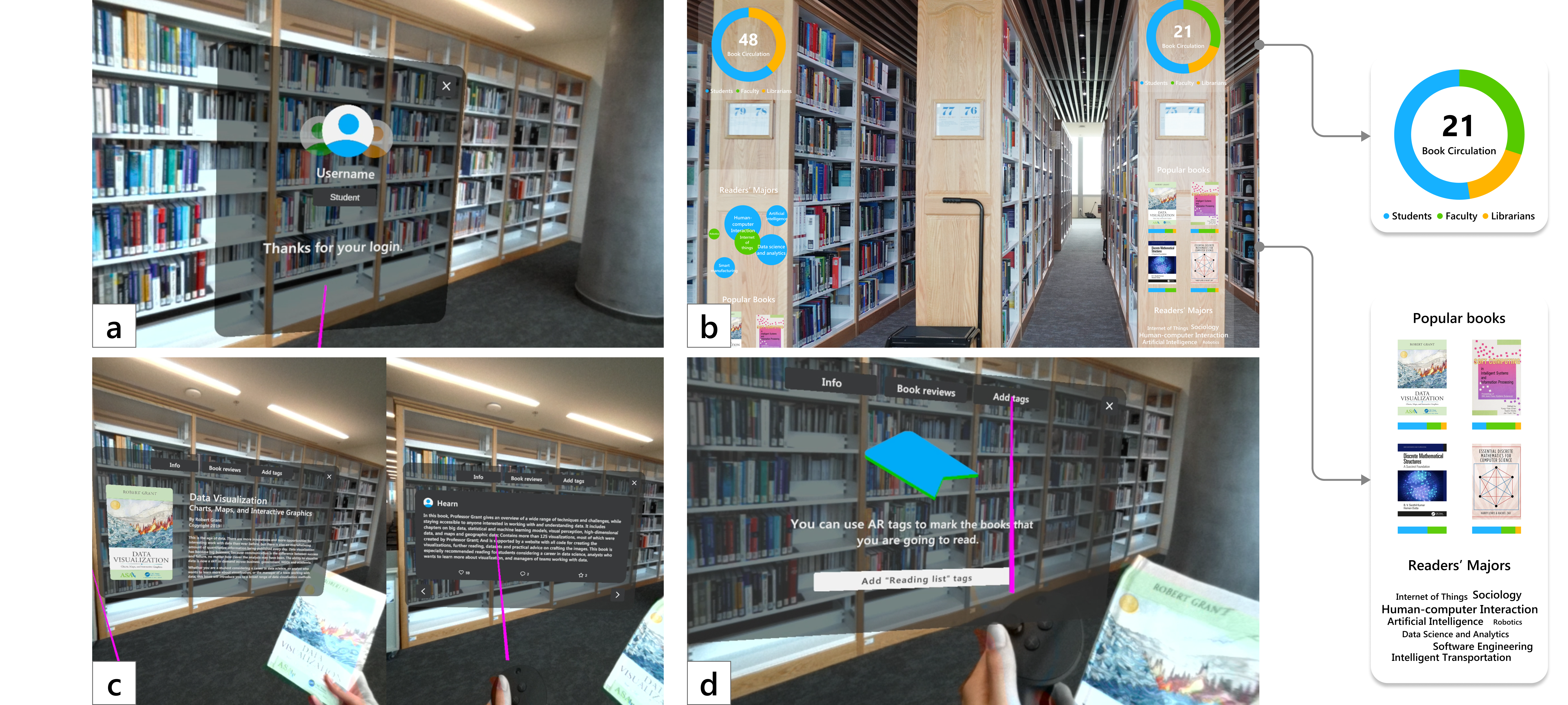}
    \caption{Key features of the AR system. (a) color-coded user groups: uses three distinct colors to distinguish between three user groups. (b) data-driven augmented bookshelves: overlays AR data visualizations on physical bookshelves. (c) interactive book reviews: users can interact with other users' reviews displayed near the physical books. (d) color-coded virtual tags: users can add virtual tags to physical books.}
    \label{fig:SystemFeatures}
\end{figure*}}

\section{Augmented Library System}

\textit{Augmented Library} incorporates several key features (\autoref{fig:SystemFeatures}) designed to enhance user engagement and interaction within the physical library environment:

\textbf{Color-coded user groups \textbf{(C3)}.}
We use three distinct colors to represent user groups associated with the library: \sethlcolor{highlightblue}\hl{blue} for students, \sethlcolor{highlightgreen}\hl{green} for faculty, and \sethlcolor{highlightyellow}\hl{yellow} for librarians. Users select their role upon logging in, and their UI elements will be displayed in corresponding color. Additionally, data visualizations on bookshelves and users' AR tags also use these colors to facilitate easy recognition among different user groups. 

\textbf{Data-driven Augmented Bookshelves \textbf{(C1)}.}
The AR content overlaying on physical bookshelves provides users with various data visualizations, including: 1) book circulation data: real-time borrowing statistics for different categories, helping users identify popular books and trends; 2) user group circulation data: borrowing habits of different user groups, enabling users to understand peer preferences; 3) popular books: highlights frequently borrowed books, encouraging users to discover noteworthy titles; 4) readers' majors' distribution: analysis of readers' academic backgrounds for specific book categories, illustrating interdisciplinary appeal.

\textbf{Interactive Book Reviews \textbf{(C2)}.}
This feature aims to enhance user engagement by visualizing book reviews as interactive AR canvas around physical books. As users approach a book, a UI canvas appears displaying various types of content, including book information and community reviews. Users can engage with reviews by liking, replying to, or favoriting them. 

\textbf{Color-coded Virtual Tags \textbf{(C3)}.}
This feature is designed to enhance the personalization of the AR library experience. Users can add AR tags to physical books, marking them for various purposes, such as "reading list." Tags are color-coded by user groups, making it easier to follow group interests (e.g., \sethlcolor{highlightblue}\hl{blue} for tags added by students). Also, faculty and librarians can add tags like "course reserve" and "new arrival," visible to all users, enhancing information sharing.

We implemented \textit{Augmented Library} in Unity (v2022.3.20f1c1) using Unity XR Interaction Toolkit and deployed it on Meta Quest Pro. Book recognition is achieved through a Reference Image Library in Unity, with the AR Tracked Images Manager in AR Foundation recognizing the books. The UI Canvas was designed following Oculus developer guidelines for comfort and attention areas.

\section{User Study}

\subsection{Participants \& Procedure}

With the IRB’s approval, we recruited 10 participants through posters in our university library community. Participants ranged in age from 22 to 34 years old (M=24.7), with 6 females and 4 males. Four had experience using AR HMDs, and most (N=6) reported infrequent library visits (once a month or less). 

The user study was conducted in a dedicated room with bookshelves, where two researchers observed and recorded participants' behavior and thoughts. The study began with a background interview. Then we used one Meta Quest Pro to deliver the AR experience. Under the guidance of researchers, participants experienced the key features of our AR system. 
Finally, we conducted semi-structured interviews with participants to gather feedback on their AR experience.
For interview data analysis, we employed thematic analysis \cite{alhojailan2012thematic}. Using open coding \cite{corbin1990grounded}, two coders independently used NVivo to code the data and discussed their decisions in weekly meetings.

\subsection{Findings}

This section presents findings from our user study, focusing on feedback and insights regarding \textit{Augmented Library}'s key features.

\subsubsection{Information Visualization and Gamified Interactions}
Interview feedback highlighted a strong preference for the AR system’s visualization capabilities, with 8 participants praising the "\textit{excellent presentation of information}" and its engaging data rendering. Some participants (N=5) suggested enhancements for book discovery, including gamified interactions. P10 suggested having a virtual pet to guide users through information to make the library experience more interesting (\autoref{fig:futureIteration}). Additionally, 4 participants (P2, P3, P6, P10) noted that the font size was too small and suggested allowing users to adjust it for better readability.

\subsubsection{Interactive Book Reviews}
\begin{figure}[htbp]
    \centering
    \includegraphics[width=\columnwidth]{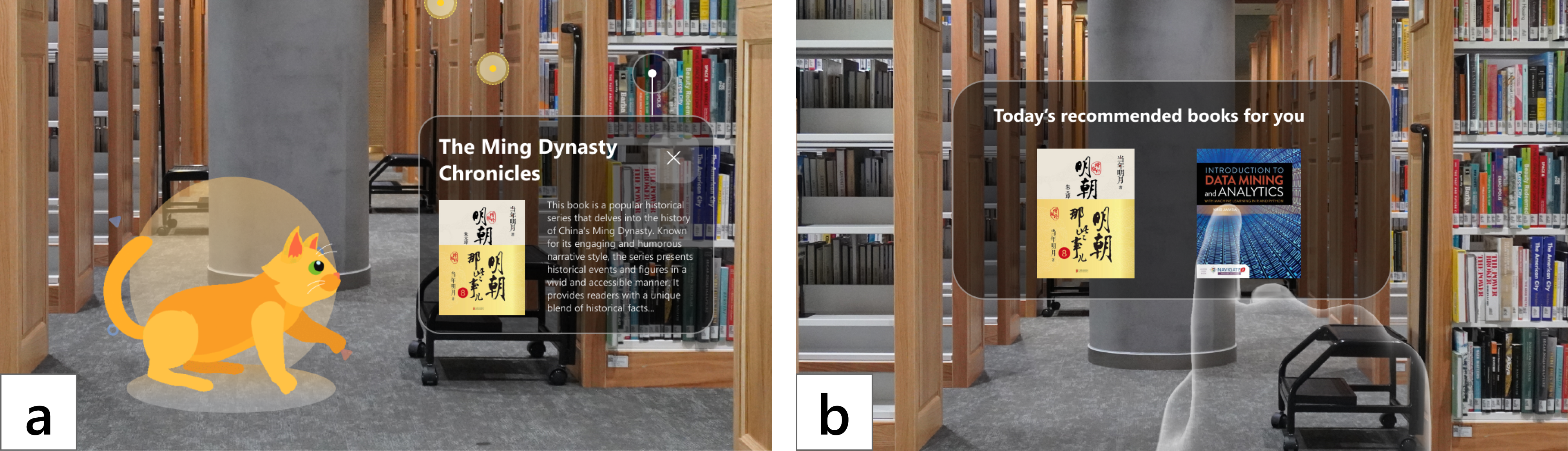}
    \caption{Participants' vision for future AR iterations. (a) a virtual cat is guiding users through information. (b) integrates AI-driven book recommendations.}
    \label{fig:futureIteration}
\end{figure}

Participants generally described the comment function as "\textit{very good}" (N=7), appreciating its ability to facilitate engagement. However, some (N=3) noted challenges in using AR to read texts, citing vision strain and physical endurance issues. P5: "\textit{I find reading the text inconvenient; it places a significant strain on my eyes.}"
Participants also emphasized the need for an AI-driven book recommendation system (\autoref{fig:futureIteration}). P8 remarked, "\textit{It's troublesome to go through reviews book by book. A recommendation system is needed to suggest books based on my requirements.}"

\subsubsection{Color-coded Virtual Tags}
Feedback on the system's tagging feature was mixed. Some participants, like P5, found that "\textit{the tagging feature makes exploration easier}," while others, such as P9, raised concerns about potential confusion.
Additionally, participants expressed the need for tag categorization. P10 proposed having both "personal" and "public" tags, noting, "\textit{There are some tags I might want only for myself, like reminders, but for interesting books, I might want to recommend them to others.}"

\section{Discussion}

In this section, we reflect on the findings from user study and discuss how our AR features address the identified challenges and contribute to enhancing user engagement within library spaces.

\textbf{Enhancing Book Discovery - C1.}
\textbf{C1} aims to improve library resource visibility through AR by overlaying data visualizations on physical bookshelves, providing real-time insights into usage trends. Positive feedback on this feature indicates AR's effectiveness in presenting complex data in an accessible and engaging manner. This aligns with previous research that suggests AR's potential in providing intuitive and real-time data visualizations \cite{chandra2019augmented,hirve2017approach,thomas2014spatial}. Our findings also showed that participants appreciated being able to see which books were popular among their peers. This informed their choices and encouraged exploration beyond their usual interests. To further enhance book discovery, future AR system iterations could consider incorporating gamification elements, which have been demonstrated as a useful way to boost engagement and motivation in educational contexts \cite{majuri2018gamification,stott2013analysis}, creating a more interactive library experience.

\textbf{Fostering Community Engagement - C2.}
We aimed to enhance the interactive and communal aspects of library through AR features like interactive book reviews, enabling users to see community insights at a glance. This approach aligns with prior work showing that social AR can facilitate information sharing and face-to-face interactions \cite{hirskyj2020social}. Our findings indicate that participants appreciated interacting with book reviews, which can not only facilitated book assessment but also sparked peer conversations. To improve this feature, they suggested simplifying the interaction process and exploring voice input \cite{zhao2020voice}, as well as integrating AI-driven recommendations for a personalized experience \cite{zhang2018towards}.

\textbf{Recognizing Diverse Stakeholders - C3.}
We considered the expectations of different library stakeholders and used color-coded user groups and virtual tags to tailor their library experience. Participants found these features helpful for recognition and interaction within the library environment, fostering better social and academic interactions, which indicates the importance of social connections in educational settings \cite{walton2012mere,wilson2013making}. However, concerns about the tagging system's complexity and potential confusion suggest that future AR iterations should streamline the interface to ensure intuitive and easy-to-use tagging features.

\section{Conclusion and Future Work}

This work explored the potential of HMD-based AR to enhance student engagement in physical libraries. Informed by a formative study with librarians, we developed \textit{Augmented Library}, featuring color-coded user groups, data-driven augmented bookshelves, interactive book reviews, and AR tags. A user study with 10 participants showed that these features enriched the library experience. However, challenges like the readability of AR content and the complexity of the virtual tagging system highlighted areas for further improvement.

Future work could focus on refining the system based on user feedback, including: enhancing AR content readability with adjustable font sizes, integrating AI-driven recommendations for personalized book discovery, streamlining the AR tagging system, and exploring gamification to boost engagement. 
Moreover, while the user study provided initial insights, its short duration limited the ability to capture comprehensive usage patterns. Future work should involve longer-term studies with an enhanced AR system incorporating user feedback and additional features.

\section{Acknowledgments}

This work is partially supported by Guangzhou-HKUST(GZ) Joint Funding Project (No.: 2024A03J0617) and HKUST Practice Research with Project title "RBM talent cultivation Exploration" (No.: HKUST(GZ)-ROP2023030).

\bibliographystyle{ACM-Reference-Format}
\bibliography{reference/1,reference/2,reference/3}

\end{document}